\documentclass[twocolumn,prl,showpacs,floatfix]{revtex4}

\usepackage{graphicx}
\usepackage{amsmath}

\begin{document}

\title
{
Nuclear dynamical deformation induced hetero- and euchromatin positioning
}

\author
{
Akinori Awazu$^{1,2}$
}

\affiliation
{
$^1$Department of Mathematical and Life Sciences, Hiroshima University,
$^2$Research Center for Mathematics on Chromatin Live Dynamics. \\
Kagami-yama 1-3-1, Higashi-Hiroshima 739-8526, Japan.
}

\date{\today}

\begin{abstract}
The contributions of active deformation dynamics in cell nuclei to the intra-nuclear positioning of hetero- and euchromatin are investigated. We analyzed the behaviors of model chains containing two types of regions, one with high and the other with low mobility, confined in a pulsating container. Here, the regions with high and low mobility represent eu- and heterochromatic regions, respectively, and the pulsating container simulates a nucleus exhibiting dynamic deformations. The Brownian dynamics simulations of this model show that the positioning of low mobility regions transition from sites near the periphery to the central region of the container if the affinity between low mobility regions and the container periphery disappears. Here, the former and latter positioning are similar to the ``conventional'' and ``inverted'' chromatin positioning observed in nuclei of normal differentiated cells and cells lacking Lamin-related proteins like mouse rod photoreceptor cell.

\end{abstract}


\maketitle

The structure and dynamics of interphase chromosomes have been recognized to play important roles in several activities in eukaryotes such as gene regulation and cell differentiation\cite{ex1,ex2,ex3,ex4,ex5,ex6,ex7,ex8,ex9,ex10,ex11,ex12}. Recently, several mesoscopic theoretical and simulation models of chromosomes were proposed to investigate mechanisms underlying the formation of several features of intra-nuclear architecture, such as the chromosome territories, transcription factories, topologically associating domains, conventional and inverted positioning of heterochromatin. \cite{model1,model2,model3,model4,model5,model6,model7,model8,model9,model10,model11}.

The cell nuclei are reportedly involved in several aspects of intra-cellular dynamics, such as the oscillation, rotation and deformation driven by the dynamics of the cytoskeleton including microtubule and actin networks\cite{move1,move2,move3,move4,move5,move6,move7,move8,move9,move10,move11,move12,move13,move14}. The dynamics of nuclei often affect the positioning and structures of intra-nuclear architectures and gene expression via interactions between the chromosomes and the nuclear periphery\cite{move1,move2,move3,move4,move5,move6,move7,move8,move9,move10,move11,move12,move14,move41,move42,move43,move44}. Moreover, the membrane proteins such as $Lamin$ $A$, $B$, $C$, and $Lamin$ receptors anchor heterochromatin to the nuclear periphery\cite{ex10,ex11,peri1,peri2,peri3,peri4,peri5,peri6}. Thus, the influence of nuclear dynamics and the physicochemical aspects of the nuclear membrane on chromosome conformations must be elucidated to clarify the activities involved in the intra-nuclear processes and their mechanisms. However, few theoretical studies have focused on such issues.

Here, we introduce a simplified chromosome model that consists of $M$ chains with two types of regions, one with high mobility and the other with low mobility, in a 3-D spherical pulsating container. The regions with high and low mobility are regarded as eu- and heterochromatic regions, respectively; heterochromatic regions consist of densely compacted DNA and proteins to silence transcription and hence, experience larger viscous drag than euchromatic regions. The pulsation of the container which is implemented by the temporal variations of the container's radius are employed to simulate the deformation of nuclei (Fig. 1(a)). Thus, by focusing on the  deformation-dependent behaviors of this model, we clarify the possible contributions of nuclear dynamics to chromatin positioning. 

In this model, each chain is constructed by $N$ spherical particles with diameter $d$ and are connected by a spring with natural length $d$. The motion of each particle is given as Brownian motion with each drug coefficient. The motion of each particle obeys

\begin{equation}
\gamma_i{\bf \dot{x}_i} = -\nabla_i (V_{int}(\{ {\bf x}_i \}) + V_{con}(\{{\bf x}_i\})) + {\bf \eta}_i(t),
\end{equation}
\begin{equation}
<{\bf \eta}_i(t){\bf \eta}_i(t')> = 2\gamma_i G \delta(t-t'),
\end{equation}
where ${\bf x}_i$ and $\gamma_i$ are the position and drug coefficient of the $i$th particle, respectively, and ${\bf \eta}_i $ and $G$ are the random force working on the $i$th particle and its magnitude, respectively. Here, $\gamma_i = \gamma_H$ for high mobility particles (H-particles) and  $\gamma_i = \gamma_L$ for low mobility particles (L-particles).

The interaction potential between particles is expressed by $V_{int}(\{ {\bf x}_i \})=V^{ch}(\{ {\bf x}_i \}) + V^{sf}(\{ {\bf x}_i \}) $, where the former represents the elastic potential between two connected particles and is given by 
\begin{equation}
V^{ch} =  \sum_{i}\frac{k_c}{2}(|{\bf x}_i-{\bf x}_{i+1}|-d)^2,
\end{equation}
and the latter represents the soft core repulsive potential due to the excluded volume between two particles; this is given by
\begin{equation}
V^{sf} = \sum_{i < j}
\begin{cases}
  \frac{k_e}{2}(|{\bf x}_i-{\bf x}_j|-d)^2 \,\,\,\,\,\, (|{\bf x}_i-{\bf x}_j| < d) \\
  0 \,\,\,\,\,\, (Otherwise)
\end{cases},
\end{equation}
with $k_e$ and $k_c$ as constants.

The motion of particle is restricted by the container; soft-core repulsion exists between the container wall and all particles, whereas short-range attraction exists between the wall and L-particles. Short-range attraction arises from the interactions between the nuclear periphery and heterochromatic regions through Lamin-related proteins. Thus, the potential is given as $\displaystyle V_{con}(\{{\bf x}_i\},R)= \sum_i V^i_{wall}({\bf x}_i,R) + \sum_{\gamma_i = \gamma_L} V^i_{lamin}({\bf x}_i,R)$, where  
\begin{equation}
V^i_{wall}({\bf x}_i, R) = 
\begin{cases}
  \frac{k_w}{2}(|{\bf x}_i|-(R-\frac{d}{2}))^2 \,\,\,\, (|{\bf x}_i| > R-\frac{d}{2}) \\
  0 \,\,\,\, (Otherwise)
\end{cases}
\end{equation}
\begin{equation}
V^i_{lamin}({\bf x}_i, R) = 
\begin{cases}
  -k_l|{\bf x}_i| \,\,\,\,\, (|{\bf x}_i| > R - d) \\
  0 \,\,\,\, (Otherwise)
\end{cases}
\end{equation}
with constants $k_w$, $k_l$ and the container radius $R$. 

We assume the container pulsates periodically by the temporally periodic variations of container's radius as   
\begin{equation}
R = R_o + A \sin \omega t
\end{equation}
where $R_o$, $A$, and $\omega$ are the basic radius, amplitude, and frequency of pulsations, respectively. 
The deformation of nuclei in cells is not periodic, but using this model, we can clarify the characteristics and mechanisms of the contributions of nuclear deformable motion to chromosome positioning. Additionally, we achieved the same qualitative results as obtained if $R$ exhibits random walk in a harmonic potential field $\frac{1}{2}k_{r}(R-R_o)^2$ with constant $k_r$. 

Now, we focus on the segregation patterns of H- and L-particles when the following chains are confined in the container: i) H-L mixed chains: each chain consists of H- and L-particle regions connected periodically, where the length of each H- and L-particle region along the center is $\lambda/2$ (see Fig. 1(b)); and ii) H-chains and L-chains:  half of all chains consist of only the H-particle region and the other half consist of only the L-particle region (see Fig. 1(c)). In the case of a large $N$, H-L mixed chains provide simple models of chromosomes containing eu- and heterochromatic regions, and H-chains and L-chains provide simple models of the euchromatin rich and heterochromatin rich chromosomes, respectively.

Here, we considered the following parameters: $\gamma_H = 1$,  $\gamma_L = 10$, $k_e = k_c = 1024$, $G = 1$, and $d = 1$. In recent experimental observations, the volume fraction of macromolecules in the nucleus is estimated as $0.2 \sim 0.3$ \cite{model10}. Thus we focused on the case of $(d/2)^3MN/R_o^3 \sim 0.296$ ($MN=512$ and $R_o=6.0$.). We confirmed the qualitative aspects of the system as shown in the bellows do not sensitive for $NM$, $R_o$, $\gamma_L$, and $\gamma_H$ when $\gamma_L > \gamma_H$ with similar volume fractions of particles. For example we obtained similar results in the case of $MN = 1024$ and $R_o=7.5d$ with $(d/2)^3MN/R_o^3 \sim 0.30$.

First, we focus on the segregation pattern of H-L mixed chains in the pulsating container. Figures 2 shows typical snapshots of the H- and L-particle distributions in two-dimensional (2-D) cross-sections (particles at $-d/2 \le x \le d/2$ on the $x-y-z$ 3-D space are shown) for $M=1$, $N=512$, $\lambda=64d$, $\omega=4\gamma_H$, and $A = d/4$ with (a) $k_l = 10$, (b) $k_l = 0$. As shown in these Figures, more H-particles (L-particles) tend to distribute near the center (periphery) of the container than H-particles (L-particles) for $k_l = 10$, while an opposite distribution occurs for $k_l = 0$. These trends in positioning of H- and L-particles are similar to the ``conventional'' and ``inverted'' eu- and heterochromatin positioning, respectively, observed in nuclei of normal differentiated cells and cells lacking Lamin-related proteins like mouse rod photoreceptor cell\cite{ex10,ex11}. 

It is easily understood that the affinity of the periphery of the container and L-particles contribute considerably to the former segregation patterns. On the other hand, if the container does not pulsate, the distributions of H- and L-particles become uniform when $k_l = 0$ because such a system is in thermodynamic equilibrium, thus indicating that the latter segregation pattern is the result of the pulsations of the container.

To observe the contributions of the pulsations of the container to pattern formations of H-L mixed chains, we focus on the pulsating frequency dependency of H- and L-particle distributions for $k_l = 0$. Figure 3 shows the relative radial distributions $P(r)$ for some $\omega$ in the cases of (a) $M=1$, $N=512$, and $\lambda=64d$, and (b) $M=4$, $N=128$, and $\lambda=64d$. Here, $P(r) = P_L(r)-P_H(r)$, where $P_m(r) = n_m(r)/4\pi r^2$ ($m =$ H or L) are the respective radial particle distributions, $r$ is the distance from the origin, and $n_m(r)$ is the frequency of $m$-particles in the region between $r$ and $r+dr$ for $dr = 0.1$. Here, $P(r) =0$ always holds when $\omega = 0$ (except for small errors due to the finite run time and sample number of simulations). The figures illustrate the appropriate values of $\omega$ for the segregation of the H- and L-particles. We obtained similar results for $\lambda = 32d$ and $\lambda = 128d$.

Similar results are obtained for H-chains and L-chains confined in the pulsating container. Figure 4 (a)-(c) show the relative radial distributions $P(r)$ for some $\omega$ in the cases of (a) $M=8$ and $N=64$ (b) $M=128$ and $N=4$, and (c) $M=512$ and $N=1$ with $k_l = 0$. As shown in these figures, the distribution of the H- and L-particles exhibits $\omega$ dependency similar to that of H-L mixed chains. However, the segregation strength becomes weak with the decrease in $N$. For example $P(d)$ that gives a typical $P(r)$ in the bulk of the container decreases drastically with the decrease in $N$ if $(dN)^{3/5}$ is smaller than $\sim (R_o-d/2)$.

The mechanisms underlying the pulsation-induced segregation pattern can be explained as follows. It is noted that H- and L-particles are equally compressed when the container contracts. On the other hand, after the expansion of the container, H-particles positioned at the outer region of particle accumulations diffuse closer to the periphery of the container faster than L-particles. Therefore, by iterating the contractions and expansions of the container, the region near the container periphery tends to be occupied by more H-particles than L-particles. 

Now, we consider the case where H-particles construct long chains such that the average distance between two edges $\sim (dN)^{3/5}$ (that can be estimated by the arguments of self-avoiding random walk\cite{poly}) is longer than $(R_o-d/2)$, or where each H-L mixed chain involves H-particle regions as long as $(\lambda/2)^{3/5} > (R_o-d/2)$. In such cases, H-particles near the center are pulled by those migrating near the periphery since they often belong to the same chain or the same region. Thus, H-particles accumulate gradually near the periphery of container. 

On the other hand, when the particles move individually or form only short chains, the influence of the pulsations on particles in the bulk of the container is negligibly weak because the amplitude of the pulsations of the container is not as large as mentioned belows. Thus, the segregation occurs only near the periphery of the container in such cases.

Notably, we found that L-particles tend to accumulate in the central region of the container in such smaller $A = d/4$ than $R_o = 6d$ for $k_l = 0$. We also confirm qualitatively the same properties of $P(r)$ in the case of $A = d/8$ although typical values of $P(d)$ become small. Thus, such segregation does not need pulsations of large amplitude. This fact indicates that weak, but not zero, active deformation of nucleus is enough for the inverted chromatin positioning in cells without Lamin-related proteins.

In this paper, we focused on the behaviors of chains containing high and low mobility regions confined in a pulsating container to understand the contributions of nuclear active deformation dynamics to the intra-nuclear positioning of hetero- and euchromatin. We found that the positioning of low mobility regions (which corresponded to heterochromatic regions) transition from sites near the periphery to the central region of the container if the affinity between these regions and the container periphery disappears. The former and latter positioning are similar to the “conventional” and “inverted” chromatin positioning observed in nuclei of normal differentiated cells and cells lacking Lamin-related proteins like mouse rod photoreceptor cell.

Recently, some theoretical models of the transition between ``conventional'' and ``inverted'' heterochromatin positioning were proposed\cite{model9,model10,model11}. These models tried to explain the mechanism of such transitions using certain assumptions (some of which could not be verified experimentally), such as the differences of chromatin-chromatin binding affinities among hetero-hetero, hetero-eu and eu-euchromatins, the difference of the characteristic length of hetero- and euchromatic regions in chromosomes. Such inhomogeneity of intra-chromosome properties may play important roles to determine the intra-nuclear chromatin positioning; however, the present results indicate that active nuclear deformation, which has been previously observed in several cells, drives the formation of ``inverted'' heterochromatin positioning without specific interactions among chromatins. Thus, we propose that the active nuclear deformations contribute considerably to the determination of intra-nuclear chromosome architectures in addition to the previously proposed inter-chromatin interactions. 

Moreover, from the point of view of polymer physics, the presented models are simple block-co-polymer populations or polymer mixtures consisting of different mobilities confined in the small space under non-equilibrium boundary conditions. Thus, more detailed theoretical studies on presented segregation pattern formation are important for soft matter and statistical physics to be performed in the future.  

The author is grateful to H. Ochiai, S. Lee, and H. Nishimori for their fruitful discussions. This research was supported in part by the Platform for Dynamic Approaches to Living System from the Ministry of Education, Culture, Sports, Science and Technology, Japan, and the Grant-in-Aid for Scientific Research on Innovative Areas ``Spying minority in biological phenomena (No.3306) (24115515)'' and ``Initiative for High-Dimensional Data-Driven Science through Deepening of Sparse Modeling (No.4503)(26120525)'' of MEXT of Japan.

\begin{figure}
\begin{center}
\includegraphics[width=8.0cm]{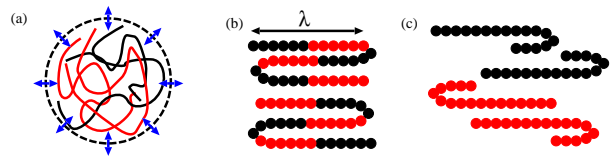}
\end{center}
\caption{(Color online) Illustrations of (a) confined chains in pulsating container, and (b) H-L mixed chains  and (c) H-chains and L-chains consisting of H-particles (grey (red)) and L-particles (black).}
\end{figure}

\begin{figure}
\begin{center}
\includegraphics[width=6.0cm]{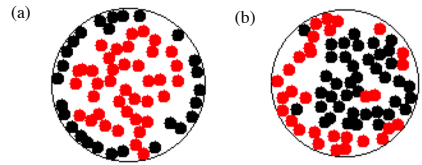}
\end{center}
\caption{(Color online) Typical snapshots of the distribution of H-particles (grey (red)) and 
L-particles (black) of H-L mixed chains ($M = 1$, $N = 512$, $\lambda = 64d$) on $x \sim 0$ 2-D cross-sections. (a) $k_l = 10$, (b) $k_l = 0$, with $\omega = 4\gamma_H$ and $A = d/4$.}
\end{figure}

\begin{figure}
\begin{center}
\includegraphics[width=8.0cm]{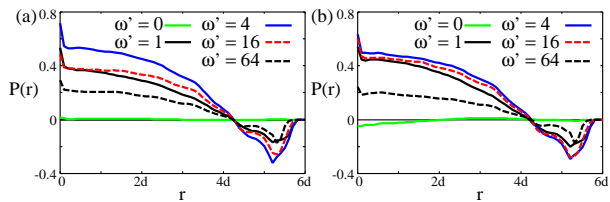}
\end{center}
\caption{(Color online) $P(r)$ of H-L mixed chains for $\omega' = \omega / \gamma_H = 0, 1, 4, 16, 64$ in the cases of (a) $M=1$, $N=512$, and $\lambda=64d$, and (b) $M=4$, $N=128$, and $\lambda=64d$ with $k_l = 0$.}
\end{figure}

\begin{figure}
\begin{center}
\includegraphics[width=8.0cm]{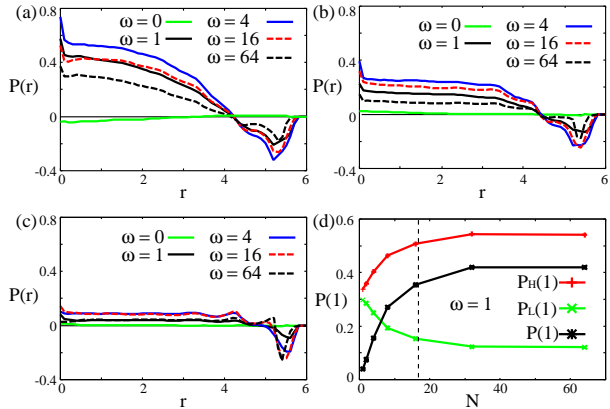}
\end{center}
\caption{(Color online) $P(r)$ of the system consists of H-chains and L-chains for $\omega' = \omega / \gamma_H = 0, 1, 4, 16, 64$ in the cases of (a) $M=8$, $N=64$, (b) $M=128$, $N=4$, and (c) $M=512$, $N=1$ with $k_l = 0$. (d) $P_H(1)$, $P_L(1)$, and $P(d)$ as function of $N$ for $NM=512$ and $\omega = 1$ with $k_l = 0$ where dot curve indicates $N = (5.5)^{5/3}$.}
\end{figure}


\begin{thebibliography}{9}

\bibitem{ex1}
T. Cremer, and C. Cremer, Nat. Rev. Genet. 2,  292 (2001).

\bibitem{ex2}
L. A. Parada, and T. Misteli , Trends Cell Biol. 12, 425 (2002).

\bibitem{ex3}
C. S. Osborne, L. Chakalova, K. E. Brown, D. Carter, A. Horton, E. Debrand, B. Goyenechea, J. A. Mitchell, S. Lopes, W. Reik, et al., Nat. Genet. 36, 1065 (2004).

\bibitem{ex4}
S. T. Kosak, and M. Groudine,  Genes Dev. 18, 1371 (2004).

\bibitem{ex5}
T. Takizawa, K. Meaburn, and T. Misteli,  Cell 135, 9 (2008).

\bibitem{ex6}
P. Meister, B. D. Towbin, B. L. Pike, A. Ponti, and S. M. Gasser,  Genes Dev. 24, 766 (2010).

\bibitem{ex7}
T. Cremer, and M. Cremer, Cold Spring Harb. Perspect. Biol. 2, a003889 (2010).

\bibitem{ex8}
P. K. Geyer, M. W. Vitalini, and L. L. Wallrath,  Curr. Opin. Cell Biol. 23, 354 (2011).

\bibitem{ex9}
W. A. Bickmore,  Annu. Rev. Genomics Hum. Genet. 14, 67 (2013).

\bibitem{ex10}
I. Solovei, M. Kreysing, C. Lanctot, S. Kosem, L. Peichl, T. Cremer, J. Guck, and B. Joffe, Cell 137, 356 (2009).

\bibitem{ex11}
I. Solovei, et al.,  Cell 152, 584 (2013).

\bibitem{ex12}
A. Bolzer, G. Kreth, I. Solovei, D. Koehler, K. Saracoglu, C. Fauth, S. Muller, R. Eils, C. Cremer, M. Speicher, et al., PLoS Biol. 3, e157 (2005).


\bibitem{model1}
M. Bohn, and D. W.  Heermann,  PLoS One 5, e12218 (2010).

\bibitem{model2}
K. Finan, P. R. Cook, and D. Marenduzzo, Chromosome Res. 19, 53 (2011).

\bibitem{model3}
M. Barbieri, M. Chotalia, J. Fraser, L. M. Lavitas, J. Dostie, A. Pombo, and M. Nicodemi,  Proc. Natl. Acad. Sci. U.S.A. 109, 16173 (2012).

\bibitem{model4}
H. Jerabek,  and D. W. Heermann,  PLoS One 7, e37525 (2012).

\bibitem{model5}
M. D. Stefano, A. Rosa, V. Belcastro, D. di Bernardo, and C. Micheletti,  PLoS Comput. Biol. 9, e1003019 (2013).

\bibitem{model6}
C. A> Brackley, S. Taylor, A. Papantonis, P. E. Cook, and D. Marenduzzo,  Proc. Natl. Acad. Sci. U.S.A. 110, 3605 (2013).

\bibitem{model7}
A. Rosa, and  R. Everaers,  Phys. Rev. Lett. 112, 118302 (2014).

\bibitem{model8}
D. Jost, P. Carrivain, G. Cavalli, and C. Vaillant,  Nucleic Acids Res. 42, 9553 (2014).


\bibitem{model9}
N. Ganai, S. Sengupta, and G. I. Menon,  Nucleic Acids Res. 42, 4145 (2014).

\bibitem{model10}
H. Jerabek, and D. W. Heermann, Int. Rev. Cell. Mol. Biol. 307, 351381 (2014).

\bibitem{model11}
A. Awazu,  Phys. Rev. E 90, 042308 (2014).


\bibitem{move1}
C. M. Pomert,  Exp. Cell Res. 5, 191 (1953).

\bibitem{move2}
K. T. S. Yao, and D. J. Ellingson,  Exp. Cell Res. 55, 39 (1969).

\bibitem{move3}
P. C. Park, and U. De Boni, Exp. Cell Res. 197, 213 (1991).

\bibitem{move4}
H. Masuda, R. Miyamoto, T. Haraguchi, and Y. Hiraoka,  Genes to Cells 11, 337 (2006).


\bibitem{move5}
K.N. Dahl, A.J. Ribeiro, and J. Lammerding, Circ. Res. 102, 1307 (2008).

\bibitem{move6}
A.C. Rowat, J. Lammerding, and J.H. Ipsen, Biophys. J. 91, 4649 (2006).

\bibitem{move7}
A. Brandt, F. Papagiannouli, J. Grosshans, et al., Curr. Biol. 16, 543 (2006).

\bibitem{move8}
A. Kumar, and G.V. Shivashankar, PLoS ONE 7, e33089 (2002).

\bibitem{move9}
D. Bhattacharya, S. Talwar, G.V. Shivashankar, et al., Biophys. J. 96, 3832 (2009).

\bibitem{move10}
J.D. Pajerowski, K.N. Dahl, D.E. Discher, et al., Proc. Natl. Acad. Sci. USA 104, 15619 (2007).

\bibitem{move11}
G.V. Shivashankar, Annu. Rev. Biophys. 40, 361 (2011).

\bibitem{move12}
S. Talwar, A. Kumar, M. Rao, G. I. Menon, and G. V. Shivashankar, Biophys. J. 104, 553 (2013).


\bibitem{move13}
K-D. Kim, et al.,  J. Cell Sci. 126, 5271 (2013).

\bibitem{move14}
N. M. Ramdas, and G. V. Shivashankar,  J. Mol. Bio. in press.

\bibitem{move41}
C. H. Thomas, J. H. Collier, C. S. Sfeir, and K. E. Healy,  Proc. Natl. Acad. Sci. U.S.A. 99, 1972 (2002).

\bibitem{move42}
K. J. Chalut, et al.,  Biophys. J. 103, 2060 (2012).

\bibitem{move43}
A. Barascu, et al.,  Nucleus 3, 411 (2012).


\bibitem{move44}
G. Fabrikant, S. Gupta, G.V. Shivashankar, and M. M. Kozlov,  Biophys. J. 105, 1316 (2013).


\bibitem{peri1}
N. Zuleger, M. I. Robson, and E. C. Schirmer. Nucleus 2, 339 (2011).

\bibitem{peri2}
L. E. Finlan, D. Sproul, I. Thomson, S. Boyle, E. Kerr, P. Perry, B. Ylstra, J. R. Chubb, and W. A. Bickmore.  PLoS Genet. 4, e1000039 (2008).

\bibitem{peri3}
J. M. Zullo, I. A. Demarco, R. Pique-Regi, D. J. Gaffney, C. B. Epstein, C. J. Spooner, T. R. Luperchio, B. E. Bernstein, J. K. Pritchard, K. L. Reddy, et al.  Cell 149, 1474 (2012).

\bibitem{peri4}
K. Van Bortle, and V. G. Corces, Cell 152, 1213 (2013).

\bibitem{peri5}
N. Zuleger, S. Boyle, D. A. Kelly, J. I. de las Heras, V. Lazou, N. Korfali, D. G. Batrakou, K. N. Randles, G. E. Morris, D. J. Harrison, et al.,  Genome Biol. 14, R14 (2013).

\bibitem{peri6}
D. Camozzi, et al.,  Nucleus 5, 427 (2014).

\bibitem{poly}
P.-G. de Gennes, Scaling Concepts in Polymer Physics (Cornell
University Press, London, 1979).



\end{thebibliography}
\end{document}